\documentclass[10pt,aps,prd,email,showpacs,preprintnumbers,showkeys,superscriptaddress,amsmath,amssymb]{revtex4-2}
\usepackage{pgfplots}
\usepackage{tikz}
\usetikzlibrary{positioning}
\usepackage{color} 
\usepackage{slashed}
\usepackage{graphicx}
\usepackage{latexsym}
\usepackage{epstopdf}
\usepackage{amsmath}
\usepackage{enumitem}
\usepackage{amssymb,amsmath}
\usepackage{multirow}
\usepackage{mathtools}
\usepackage{bbm}
\usepackage{bm}
\usepackage{amsfonts}
\usepackage{amssymb}
\usepackage{tabularx}
\usepackage{calligra}
\usepackage{calrsfs}
\usepackage{makecell}
\usepackage[normalem]{ulem}
\usepackage[USenglish,american]{babel}
\usepackage{xcolor}
\usepackage{hyperref}
\hypersetup{
  colorlinks=false,             
  linkbordercolor=blue,         
  citebordercolor=green,
  urlbordercolor=blue,
  pdfborder={0 0 1}             
}
\usepackage{subcaption}
\usepackage[compat=1.1.0]{tikz-feynman}

\expandafter\ifx\csname package@font\endcsname\relax\else
 \expandafter\expandafter
 \expandafter\usepackage
 \expandafter\expandafter
 \expandafter{\csname package@font\endcsname}%
\fi
\begin{document}
\title{Gauge-covariant stochastic neural fields: Stability and finite-width effects}

\author{Rodrigo Carmo Terin} 
\email{rodrigo.carmo@urjc.es}
\affiliation{King Juan Carlos University, Facuty of Experimental Sciences and Technology, Department of Applied Physics, Av. del Alcalde de Móstoles, 28933, Madrid, Spain}

\begin{abstract}
We develop a gauge-covariant stochastic effective field theory for stability and finite-width effects in deep neural systems. The model uses classical commuting fields: a complex matter field, a real Abelian connection field, and a fictitious stochastic depth variable. Using the Martin--Siggia--Rose--Janssen--de~Dominicis formalism, we derive its functional representation and a two-replica linear-response construction defining the maximal Lyapunov exponent and the amplification factor for the edge of chaos. Finite-width effects appear as perturbative corrections to dressed kernels, and the marginality condition remains unchanged at the order considered for fixed kernel geometry. Numerically, finite-width multilayer perceptrons follow the mean-field instability threshold, and a linear stochastic effective sector reproduces the predicted low-frequency spectral deformation.
\end{abstract}



\maketitle

\section{Introduction}
\label{sec_1}

Deep neural networks have achieved remarkable success in domains such as computer vision~\cite{krizhevsky2012imagenet,he2015deep}, speech recognition~\cite{hannun2014deepspeech,sak2014lstm}, and natural language processing~\cite{vaswani2017attention,devlin2018bert}. Despite this empirical success, the theoretical principles governing stability, information propagation, and the onset of instability in deep architectures remain only partially understood~\cite{hutson2018alchemy,roberts2021why,roberts2021principles}. In practice, activation functions, initialization scales, normalization schemes, and architectural constraints are often selected by a mixture of theory and heuristics, especially near the so-called edge of chaos, where perturbations neither vanish too rapidly nor explode across depth~\cite{Sompolinsky1988,poole2016exponential,Langton1990,Bertschinger2004,Laje2013}. One of the most fruitful directions in recent years has been the development of correspondences between neural networks and statistical or quantum field theory. In the infinite-width limit, many architectures admit Gaussian-process or kernel descriptions~\cite{neal1995bayesian,lee2018deep}, while finite-width effects can be organized as controlled corrections to these large-width limits~\cite{yaida2020nonGaussian,dyer2020asymptotics,halverson2021neural,erbin2022nonperturbative}. This has led to a broader neural-network--field-theory perspective in which propagators, generating functionals, saddle points, and diagrammatic expansions provide useful tools for understanding architecture-dependent behavior~\cite{grosvenor2022edge,ferko2025quantum,halverson2024conformal,roberts2021principles}.

Most works in this direction, however, are based on global symmetries, large-\(N\) vector models, or field-theoretic descriptions without an explicit local gauge structure~\cite{brezin1973critical,moshe2003quantum,halverson2021neural,erbin2022nonperturbative,ferko2025quantum,halverson2024conformal}. By contrast, local symmetries are among the most powerful organizing principles in theoretical physics: they constrain admissible interactions, control longitudinal sectors, and generate functional identities that strongly restrict perturbative corrections. This naturally raises the question of whether a local gauge-covariant structure can also serve as a useful organizing principle for neural stability and finite-width dynamics.
This question is additionally motivated by recent efforts to incorporate gauge ideas directly into machine-learning architectures. Explicit gauge-invariant or gauge-equivariant neural models have already been explored in several contexts~\cite{maiti2021symmetry,bondesan2021hintons,theodosis2024gauge,Luo:2022jzl}. These approaches typically enforce local symmetry at the architectural or feature-engineering level. The goal of the present work is different but complementary: rather than proposing a single microscopic architecture, we develop a continuum stochastic effective theory with local \(U(1)\) covariance and use it to analyze linear-response stability, marginality, and finite-width corrections in deep neural propagation.

Here, we do not claim a literal equivalence between neural networks and quantum electrodynamics. Nor do we formulate neural activations as Grassmann-valued fermions. Instead, we introduce a mathematically consistent effective stochastic field theory built from classical commuting fields: a complex matter field representing coarse-grained feature amplitudes, a real Abelian connection field representing effective connectivity structure, and a fictitious stochastic depth variable governing noisy propagation. The resulting framework is gauge-theoretic in structure, but it remains an effective classical stochastic model throughout.

More concretely, we consider a complex field \(\phi(x,t)\), its conjugate \(\phi^\ast(x,t)\), and a real field \(W_\mu(x,t)\), where \(x\) is an effective coordinate labeling feature, spatial, or latent directions depending on architecture, and \(t\) is a fictitious stochastic depth or Langevin variable. The local \(U(1)\) transformations
\begin{eqnarray}
\phi(x,t) \to e^{i\theta(x,t)}\phi(x,t),\qquad
\phi^\ast(x,t) \to e^{-i\theta(x,t)}\phi^\ast(x,t),\qquad
W_\mu(x,t) \to W_\mu(x,t)-\frac{1}{g}\partial_\mu\theta(x,t)\nonumber\\
\end{eqnarray}
define the covariance structure of the model. The effective action is constructed from the corresponding covariant derivative and field strength, and the noisy depth evolution is formulated through It\^o Langevin equations and their Martin--Siggia--Rose--Janssen--de~Dominicis (MSRJD) functional representation~\cite{martinsiggia1973path,janssen1976lagrangian,dedom1975lagrangian,Ito1944,Ito1951}.
In this framework, the analogy with Abelian gauge theory is structural rather than literal. The formal comparison with Euclidean gauge theory is useful because it provides a disciplined language for local covariance, gauge fixing, longitudinal/transverse decomposition, Ward-type identities, and perturbative dressing of kernels. At the same time, the neural interpretation remains distinct: the coordinate \(x\) is an effective modeling coordinate rather than physical spacetime, the parameter \(d\) is an effective dimension rather than a fundamental one, and the gauge parameter \(\alpha\) may also be interpreted as labeling a family of effective kernel geometries on the neural side.

This clarification is important because dimensional regularization and Euclidean field-theory notation appear naturally in the effective continuum treatment. In the present work, the use of \(d=4\) should be understood as a convenient benchmark inherited from Abelian gauge-theory notation, not as an intrinsic property of the neural system. Likewise, dimensional continuation \(d=4-\epsilon\) is employed only as a regularization device for the effective continuum kernels. Nothing in the neural interpretation requires a physical four-dimensional spacetime.
The main problem addressed in this paper is the stability of perturbation growth across stochastic depth. To that end, we formulate a two-replica linear-response construction, in which two copies of the same stochastic system evolve under the same noise realization but start from slightly different initial conditions. This makes it possible to define the maximal Lyapunov exponent \(\lambda_{\max}\) and, from the dominant dressed fluctuation channel, a full amplification factor \(\chi\). In the revised presentation, the edge of chaos is identified with the marginality condition \(\lambda_{\max}=0\), equivalently \(\chi=1\), where \(\chi\) always denotes the full dressed-to-bare gain rather than only its perturbative correction.

A second goal of the paper is to understand finite-width effects in the same language. In wide-network theory, finite-width corrections modify propagation away from the mean-field limit~\cite{yaida2020nonGaussian,dyer2020asymptotics,erbin2022nonperturbative}. In the present framework, these corrections appear as perturbative deformations of the effective response and correlation kernels. Local covariance constrains the allowed structure of such corrections through Ward-type identities. The practical consequence is not that every observable becomes universal, but that specific perturbative corrections can renormalize amplitudes and spectral weights without shifting the marginality condition within a fixed kernel geometry at the perturbative order considered.
To test these ideas numerically, we perform two complementary studies. First, we analyze finite-width multilayer perceptrons at initialization and compare the empirical perturbation-growth exponent with the corresponding mean-field amplification criterion for \(\tanh\) and ReLU activations. Second, we study a linear stochastic effective model as a controlled toy sector in which the leading finite-width correction to the spectral kernel can be compared directly against simulation in frequency space. These numerical experiments are intentionally interpreted conservatively: they validate the stability logic and the effective dressed-kernel description, but they are not presented as literal numerical realizations of quantum electrodynamics.

The present work is also connected to earlier developments in stochastic quantization and gauge theory~\cite{parisi1981stochastic,faddeev1967,peskin2018introduction,marciano1978,ball1980,karplus1950,karplus1951,loveridge2021a,loveridge2021b,loveridge2022,oliveira2019,kizilersu2015,bashir2014,gies2012}. We emphasize, however, that these references play a methodological rather than ontological role here. They provide the mathematical and conceptual background for local covariance, gauge fixing, and dressed-kernel reasoning, but the model studied in this paper is a classical stochastic effective theory designed for neural stability analysis. Taken in this sense, the paper contributes to the broader program of importing controlled symmetry-based tools from field theory into machine learning, while avoiding stronger equivalence claims than the formalism can justify. The main contributions are as follows:
\begin{enumerate}[label=\roman*.]
    \item We formulate a gauge-covariant stochastic effective field theory for neural propagation using only commuting fields, thereby removing the ambiguity of a fermionic analogy in the neural matter sector.
    \item We derive the MSRJD functional representation and use a two-replica linear-response construction to define the maximal Lyapunov exponent and the full amplification factor governing the edge-of-chaos condition.
    \item We show how finite-width effects can be organized as perturbative corrections to dressed kernels and explain, in a precise perturbative sense, why the marginality condition is not shifted at the order considered within a fixed kernel geometry.
    \item We support the framework with numerical evidence from finite-width multilayer perceptrons and from a linear stochastic effective sector whose spectral deformation can be compared directly with the leading perturbative prediction.
\end{enumerate}
This paper is organized as follows. Section~\ref{sec:effective_model} introduces the gauge-covariant stochastic effective model and clarifies the meaning of the effective coordinates, fields, and local symmetry. Section~\ref{sec_4} derives the MSRJD representation and the relevant response and correlation observables. Section~\ref{sec:structural_comparison} explains the precise structural relation with Abelian gauge theory and clarifies the role of the effective dimension \(d\). Section~\ref{sec:stability_edge_chaos} develops the two-replica stability analysis, defines the amplification factor, and formulates the edge-of-chaos criterion. Section~\ref{sec:numerical_results} presents the numerical results. Section~\ref{sec:local_symmetry_kernel} discusses local symmetry, kernel geometry, and admissible observables. Section~\ref{sec:conclusion} concludes. Appendix~\ref{app:msrjd_derivation} contains the MSRJD derivation, Appendix~\ref{app:ward_identity} derives the Ward-type identity, Appendix~\ref{app:oneloop} discusses the leading perturbative correction to the amplification kernel, and Appendix~\ref{app:finite_width_linear} summarizes the finite-width expansion in the linear stochastic effective sector.


\section{Effective gauge-covariant stochastic field model for neural dynamics}
\label{sec:effective_model}

In this section we introduce the effective stochastic field model that underlies the rest of the paper. 
The purpose of the construction is not to identify a neural network with quantum electrodynamics in a literal sense, 
but to endow neural dynamics with a controlled local \(U(1)\) gauge-covariant structure that can be analyzed with field-theoretic tools.
Throughout this work, all dynamical variables are \emph{classical commuting fields}. 
In particular, the matter sector is described by a complex scalar field and not by Grassmann-valued fermions. 
This point is essential: the revised formulation is an effective stochastic gauge theory for neural dynamics, inspired by Abelian gauge theory, but not a fermionic QED path integral.

We consider an effective field description in terms of:
\begin{itemize}
    \item a complex matter field \(\phi(x,t)\in\mathbb{C}\), representing coarse-grained neural activations or feature amplitudes;
    \item its complex conjugate \(\phi^\ast(x,t)\);
    \item a real Abelian gauge field \(W_\mu(x,t)\), representing an effective connectivity or phase-transport field;
    \item a fictitious evolution variable \(t\), interpreted as continuous depth or Langevin time;
    \item an effective coordinate \(x\in\Omega\subset\mathbb{R}^d\), representing feature space, spatial position, or another architecture-dependent latent coordinate.
\end{itemize}

The coordinate \(x\) should therefore be understood as an \emph{effective modeling coordinate}, not as physical spacetime. 
Its meaning depends on the architecture under consideration:
for multilayer perceptrons it can encode feature channels or neuron groups;
for convolutional architectures it may represent spatial position together with channel structure;
for graph-based models it can be regarded as a continuum proxy for node embeddings or graph coordinates.
The dimension \(d\) is likewise an effective modeling parameter. 
It is \emph{not} fixed by any physical requirement of the neural network.
Whenever \(d=4\) is used below, it should be understood as a convenient benchmark dimension inherited from standard gauge-theory notation, not as an intrinsic property of the learning system.
All formal manipulations can be carried out in general \(d\), and dimensional continuation is used only as an analytical regularization device.

The basic local symmetry is
\begin{eqnarray}
\phi(x,t) &\to& e^{i\theta(x,t)}\,\phi(x,t),\\
\phi^\ast(x,t) &\to& e^{-i\theta(x,t)}\,\phi^\ast(x,t),\\
W_\mu(x,t) &\to& W_\mu(x,t)-\frac{1}{g}\partial_\mu\theta(x,t),
\end{eqnarray}
where \(g\) is a coupling parameter and \(\theta(x,t)\) is an arbitrary local phase.
The corresponding covariant derivative is
\begin{eqnarray}
D_\mu \phi := (\partial_\mu + i g W_\mu)\phi,
\qquad
(D_\mu \phi)^\ast = (\partial_\mu - i g W_\mu)\phi^\ast,
\end{eqnarray}
and the field strength is
\begin{eqnarray}
F_{\mu\nu}=\partial_\mu W_\nu-\partial_\nu W_\mu.
\end{eqnarray}
This local symmetry is interpreted on the neural side as a redundancy in the phase-like parametrization of intermediate feature variables, compensated by a transformation of the effective connectivity field.
The role of gauge covariance here is therefore organizational and structural: it constrains the admissible effective dynamics and the class of observables that are meaningful in the coarse-grained description.

The Euclidean effective action is chosen as
\begin{eqnarray}
S_{\rm eff}[\phi^\ast,\phi,W]
=
\int d^d x \,
\bigg[
(D_\mu\phi)^\ast (D_\mu\phi)
+ m^2 \phi^\ast\phi
+ U(\phi^\ast\phi)
+ \frac{1}{4}F_{\mu\nu}F_{\mu\nu}
+ \frac{1}{2\alpha}(\partial_\mu W_\mu)^2
\bigg],\nonumber\\
\label{eq:Seff_main}
\end{eqnarray}
where:
\begin{itemize}
    \item \(m\) is an effective mass or regularization scale;
    \item \(U(\phi^\ast\phi)\) is a local gauge-invariant potential;
    \item \(\alpha\) is a gauge-fixing parameter in the field-theoretic representation.
\end{itemize}
The term \((\partial_\mu W_\mu)^2/(2\alpha)\) should be interpreted carefully.
In the gauge-theory language it is the standard covariant gauge-fixing term.
In the neural interpretation it parametrizes a family of effective kernel geometries or propagation metrics in the connectivity sector.
Thus, changing \(\alpha\) does not merely correspond to a redundancy of description on the neural side; rather, it labels different effective kernels within the same gauge-covariant modeling class.
The action in Eq.~\eqref{eq:Seff_main} is the central object of the revised framework.
It is mathematically well-defined for commuting complex fields and avoids the inconsistencies that arise when fermionic expressions are transplanted directly into a commuting setting.

To model noisy propagation and finite-width fluctuations, we introduce a fictitious stochastic evolution in \(t\).
The fields obey It\^o Langevin equations of the form
\begin{eqnarray}
\partial_t \phi(x,t)
&=&
-\,\frac{\delta S_{\rm eff}}{\delta \phi^\ast(x,t)}
+\eta(x,t),
\label{eq:langevin_phi}
\\
\partial_t \phi^\ast(x,t)
&=&
-\,\frac{\delta S_{\rm eff}}{\delta \phi(x,t)}
+\eta^\ast(x,t),
\label{eq:langevin_phistar}
\\
\partial_t W_\mu(x,t)
&=&
-\,\frac{\delta S_{\rm eff}}{\delta W_\mu(x,t)}
+\xi_\mu(x,t).
\label{eq:langevin_W}
\end{eqnarray}
Here \(\eta(x,t)\) is a complex Gaussian white noise, \(\eta^\ast(x,t)\) its complex conjugate, and \(\xi_\mu(x,t)\) is a real Gaussian white noise.
Their nonvanishing correlations are taken as
\begin{eqnarray}
\langle \eta(x,t)\eta^\ast(x',t')\rangle
&=&
2\kappa_\phi\,
\delta^{(d)}(x-x')\delta(t-t'),
\\
\langle \xi_\mu(x,t)\xi_\nu(x',t')\rangle
&=&
2\kappa_W\,
\delta_{\mu\nu}\delta^{(d)}(x-x')\delta(t-t').
\end{eqnarray}
The two noise amplitudes \(\kappa_\phi\) and \(\kappa_W\) control, respectively, stochasticity in the matter and connectivity sectors.
In the neural interpretation they encode effective fluctuations associated with activation noise, random updates, finite-width variability, or stochastic optimization effects.
For notational simplicity one may later specialize to \(\kappa_\phi=\kappa_W\equiv\kappa\), but no conceptual step depends on this simplification.

\subsection{Why this model is appropriate}

This formulation has four advantages relevant to the present work.
First, it is formally consistent: all fields are commuting and the stochastic measure is defined with ordinary Gaussian noises.
Second, it preserves the local \(U(1)\) structure needed to derive Ward-type identities and constrain longitudinal fluctuations.
Third, it admits a direct MSRJD representation, which makes perturbative corrections and response functions accessible in a controlled way.
Fourth, it provides a natural effective language for neural stability analysis:
the matter sector describes propagation of feature amplitudes,
the gauge sector describes structured couplings or transport of phase information,
and the stochastic evolution in \(t\) captures noisy depth dynamics and finite-width effects.
For these reasons, Eq.~\eqref{eq:Seff_main} will be our starting point throughout the rest of the paper.
The next section reformulates the Langevin system in the Martin--Siggia--Rose--Janssen--de~Dominicis language and identifies the observables relevant for stability and edge-of-chaos analysis.


\section{MSRJD formulation and observables for stochastic neural stability}
\label{sec:msrjd_observables}

The stochastic evolution introduced in Sec.~\ref{sec:effective_model} admits a Martin--Siggia--Rose--Janssen--de~Dominicis (MSRJD) representation, which provides the functional framework used throughout this work to compute response functions, correlation functions, and stability diagnostics. In the present context, the MSRJD construction is not an auxiliary analogy but the precise stochastic field-theoretic formulation of the effective neural model itself.
Starting from the It\^o Langevin equations
\begin{eqnarray}
\partial_t \phi(x,t)
&=&
-\,\frac{\delta S_{\rm eff}}{\delta \phi^\ast(x,t)}+\eta(x,t),
\\
\partial_t \phi^\ast(x,t)
&=&
-\,\frac{\delta S_{\rm eff}}{\delta \phi(x,t)}+\eta^\ast(x,t),
\\
\partial_t W_\mu(x,t)
&=&
-\,\frac{\delta S_{\rm eff}}{\delta W_\mu(x,t)}+\xi_\mu(x,t),
\end{eqnarray}
we introduce response fields \(\tilde{\phi}^\ast(x,t)\), \(\tilde{\phi}(x,t)\), and \(\tilde{W}_\mu(x,t)\), which impose the stochastic equations through functional delta constraints. After averaging over the Gaussian noises, the generating weight becomes
\begin{eqnarray}
Z
&=&
\int \mathcal{D}\phi\,\mathcal{D}\phi^\ast\,\mathcal{D}W\,
\mathcal{D}\tilde{\phi}\,\mathcal{D}\tilde{\phi}^\ast\,\mathcal{D}\tilde{W}\;
e^{-S_{\rm MSRJD}},
\end{eqnarray}
with MSRJD action
\begin{eqnarray}
S_{\rm MSRJD}
&=&
\int dt\,d^dx\;\Bigg\{
\tilde{\phi}^\ast\!\left[\partial_t\phi+\frac{\delta S_{\rm eff}}{\delta \phi^\ast}\right]
+
\tilde{\phi}\!\left[\partial_t\phi^\ast+\frac{\delta S_{\rm eff}}{\delta \phi}\right]
+
\tilde{W}_\mu\!\left[\partial_t W_\mu+\frac{\delta S_{\rm eff}}{\delta W_\mu}\right]
\nonumber\\
&&
\qquad\qquad
-\kappa_\phi\,\tilde{\phi}^\ast\tilde{\phi}
-\kappa_W\,\tilde{W}_\mu\tilde{W}_\mu
\Bigg\}.
\label{eq:MSRJD_main}
\end{eqnarray}
This expression is well-defined entirely in terms of commuting fields and ordinary Gaussian noises. In particular, no Grassmann-valued neural variables or fermionic stochastic sources are required anywhere in the construction. This point removes the ambiguity present in the original formulation and makes the stochastic measure mathematically consistent from the outset.

The response fields have the standard dynamical meaning of MSRJD theory: they measure the sensitivity of the stochastic evolution to perturbations in the drift terms and external sources. In the neural interpretation, they play the role of linear-response variables associated with how perturbations in effective activations and connectivity propagate along the fictitious depth variable \(t\).
More concretely, while \(\phi\) and \(W_\mu\) encode the forward stochastic evolution of feature amplitudes and effective couplings, the response fields encode how these quantities react to infinitesimal disturbances. This is the precise functional origin of the stability observables used later in the edge-of-chaos analysis.

To generate correlation and response functions, we couple external sources to both dynamical and response fields:
\begin{eqnarray}
Z[J,J^\ast,J_\mu;\tilde{J},\tilde{J}^\ast,\tilde{J}_\mu]
&=&
\int \mathcal{D}\Phi\,\mathcal{D}\tilde{\Phi}\;
\exp\Bigg[
- S_{\rm MSRJD}
+ \int dt\,d^dx\,
\Big(
J^\ast \phi + J \phi^\ast + J_\mu W_\mu
\nonumber\\
&&
\qquad\qquad\qquad\qquad
+ \tilde{J}^\ast \tilde{\phi} + \tilde{J}\tilde{\phi}^\ast + \tilde{J}_\mu \tilde{W}_\mu
\Big)
\Bigg],
\end{eqnarray}
where
\begin{eqnarray}
\Phi=(\phi,\phi^\ast,W_\mu),\qquad
\tilde{\Phi}=(\tilde{\phi},\tilde{\phi}^\ast,\tilde{W}_\mu).
\end{eqnarray}
Functional derivatives of \(Z\), or equivalently of the connected generator \(W=\ln Z\), produce the observables of interest. The effective action obtained by Legendre transform can then be used to organize perturbative corrections around mean-field solutions in the usual way.

The fundamental observables for the present work are the two-point correlation and response functions. In particular, we define the matter correlator
\begin{eqnarray}
C_{\phi\phi^\ast}(x,t;x',t')
&:=&
\big\langle \phi(x,t)\phi^\ast(x',t')\big\rangle,
\end{eqnarray}
the connectivity correlator
\begin{eqnarray}
C_{\mu\nu}(x,t;x',t')
&:=&
\big\langle W_\mu(x,t)W_\nu(x',t')\big\rangle,
\end{eqnarray}
and the corresponding linear response functions
\begin{eqnarray}
R_{\phi\phi^\ast}(x,t;x',t')
&:=&
\big\langle \phi(x,t)\tilde{\phi}(x',t')\big\rangle,
\\
R_{\mu\nu}(x,t;x',t')
&:=&
\big\langle W_\mu(x,t)\tilde{W}_\nu(x',t')\big\rangle.
\end{eqnarray}
These objects control the propagation of fluctuations and perturbations across the effective depth dynamics. In particular, the long-\(t\) growth or decay of suitable linearized combinations of these correlators determines whether the system lies in an ordered, marginal, or unstable regime.
At the quadratic level, the MSRJD action defines the bare propagators of the effective theory. Higher-order contributions generated by the interaction terms in \(S_{\rm eff}\) then produce finite-width or finite-fluctuation corrections in direct analogy with standard loop expansions in stochastic field theory. What matters for the present paper is that these corrections are computed inside a well-defined bosonic stochastic model, rather than by importing fermionic expressions unchanged from QED.

Because the effective dynamics is built from the gauge-covariant action \(S_{\rm eff}\), only quantities with a clear meaning under local \(U(1)\) transformations should be regarded as physically meaningful observables of the coarse-grained neural model. This includes, for example, gauge-invariant composites such as \(|\phi|^2=\phi^\ast\phi\), correlation functions defined after gauge fixing, and response-based stability indicators evaluated at fixed kernel geometry.
An important distinction must be kept in mind. In an ordinary gauge theory, changing the gauge parameter \(\alpha\) is only a change of description. In the neural interpretation adopted here, however, \(\alpha\) also labels different effective propagation kernels within the model class. Therefore, observables are invariant under local \(U(1)\) reparametrizations at fixed \(\alpha\), but they need not remain numerically unchanged when \(\alpha\) itself is varied. This distinction will be central in the interpretation of stability thresholds below.

The MSRJD formulation provides the technical backbone for the rest of the manuscript. First, it gives a controlled mean-field limit by saddle-point analysis of the stochastic functional. Second, it identifies the response sector needed for Lyapunov and edge-of-chaos diagnostics. Third, it furnishes the perturbative language in which finite-width corrections can be organized and interpreted.
For this reason, the next section does not postulate an exact identification with QED, but instead introduces a more careful structural comparison with Abelian gauge theory, emphasizing what is genuinely shared by the two frameworks and what remains specific to the neural effective model developed here.



\section{Structural comparison with Abelian gauge theory}
\label{sec:structural_comparison}

Having established the effective stochastic model and its MSRJD representation, we now clarify in what precise sense the present framework is related to Abelian gauge theory. The comparison is structural rather than literal: the neural model developed in this work is a classical stochastic field theory with local \(U(1)\) covariance, while Euclidean Abelian gauge theory is a quantum field theory whose standard matter sector may be fermionic or bosonic depending on the physical model under consideration.
The purpose of the comparison is therefore not to claim an exact identification with QED, but to isolate the mathematical ingredients that are genuinely shared by both descriptions: local phase covariance, covariant derivatives, gauge-field propagation, gauge fixing, Ward-type constraints, and perturbative organization of fluctuations.

At the level of field content and local symmetry, the effective neural model of Sec.~\ref{sec:effective_model} has the same basic Abelian gauge-theory architecture as a Euclidean charged matter field minimally coupled to a \(U(1)\) connection. In both cases one has:
\begin{itemize}
    \item a matter field carrying a local phase;
    \item a gauge connection entering through a covariant derivative;
    \item a field-strength tensor \(F_{\mu\nu}\);
    \item a gauge-fixing sector parametrized by \(\alpha\);
    \item correlation and response functions constrained by local symmetry.
\end{itemize}
For the present model, the relevant Euclidean action is
\begin{eqnarray}
S_{\rm eff}[\phi^\ast,\phi,W]
=
\int d^dx \,
\bigg[
(D_\mu\phi)^\ast (D_\mu\phi)
+ m^2 \phi^\ast\phi
+ U(\phi^\ast\phi)
+ \frac{1}{4}F_{\mu\nu}F_{\mu\nu}
+ \frac{1}{2\alpha}(\partial_\mu W_\mu)^2
\bigg],\nonumber\\
\end{eqnarray}
which should be compared with the standard Euclidean action of an Abelian gauge theory with complex matter field \( \varphi \):
\begin{eqnarray}
S_{\rm Abelian}[\varphi^\ast,\varphi,A]
=
\int d^dx \,
\bigg[
(\mathcal{D}_\mu\varphi)^\ast (\mathcal{D}_\mu\varphi)
+ m^2 \varphi^\ast\varphi
+ V(\varphi^\ast\varphi)
+ \frac{1}{4}F_{\mu\nu}[A]F_{\mu\nu}[A]
+ \frac{1}{2\alpha}(\partial_\mu A_\mu)^2
\bigg],\nonumber\\
\label{eq:abelian_reference_action}
\end{eqnarray}
with \(\mathcal{D}_\mu=\partial_\mu+ieA_\mu\).
Under the structural dictionary
\begin{eqnarray}
\phi \leftrightarrow \varphi,\qquad
W_\mu \leftrightarrow A_\mu,\qquad
g \leftrightarrow e,
\end{eqnarray}
the effective neural action and the Abelian gauge-theory action have the same gauge-covariant form. This is the precise level at which the correspondence should be understood in the revised manuscript.
The shared structure is substantial but limited. The following ingredients are genuinely common to both frameworks:
\begin{enumerate}
    \item local \(U(1)\) covariance and the associated transformation laws;
    \item the use of covariant derivatives and gauge-field strengths;
    \item gauge fixing and longitudinal/transverse decomposition;
    \item the existence of Ward-type identities derived from local symmetry;
    \item the possibility of organizing corrections through perturbative expansions around quadratic kernels.
\end{enumerate}
By contrast, the following aspects are not being identified:
\begin{enumerate}
    \item the neural model is not a quantum theory;
    \item the fields \(\phi,\phi^\ast,W_\mu\) are not particle operators or quantum amplitudes;
    \item the coordinate \(x\) is an effective feature-space coordinate, not physical spacetime;
    \item the stochastic variable \(t\) is an auxiliary depth or Langevin coordinate, not physical time;
    \item the parameter \(\alpha\) plays an interpretive role on the neural side that differs from pure gauge redundancy in particle physics.
\end{enumerate}
These distinctions are essential for avoiding overstatement. In particular, nothing in the present paper requires a literal QED interpretation of the neural model, and no claim of physical equivalence between deep networks and elementary matter fields is made.

One of the main points requiring clarification in the original submission concerns the role of the coordinate \(x\) and the dimension \(d\). In the revised formulation, \(x\in\mathbb{R}^d\) is an effective coordinate used to organize the geometry of the coarse-grained feature representation. The value of \(d\) depends on the intended architecture-level interpretation and is not fixed by any fundamental principle.

For multilayer perceptrons, \(x\) may represent a continuous proxy for feature indices or neuron groups. For convolutional architectures, it may encode spatial position and channels. For graph-based models, it may represent an embedding coordinate or a continuum approximation to graph locality. In all cases, \(d\) is a modeling choice, not a physical spacetime dimension.
The use of \(d=4\) in some formulae is therefore only a convenient benchmark inherited from standard gauge-theory notation. It is not essential to the neural interpretation of the model. Likewise, dimensional continuation \(d=4-\epsilon\) is employed purely as an analytical regularization device. On the neural side, this continuation should be understood as a formal way of regulating momentum integrals in the effective continuum theory, not as a statement about fractional physical dimensions of a network.

In Abelian gauge theory, the gauge-fixing parameter \(\alpha\) labels different descriptions of the same physical content. In the present neural effective model, the situation is subtler. The parameter \(\alpha\) still enters through the same quadratic gauge-field kernel, but on the neural side this kernel also characterizes an effective propagation geometry for the connectivity sector.

This means that changing \(\alpha\) can be read in two different ways depending on context. From the gauge-theory viewpoint it is a choice of gauge representation. From the neural viewpoint it may also be interpreted as varying the effective kernel geometry within the same gauge-covariant modeling class. Accordingly, statements about invariance must always be interpreted at fixed \(\alpha\) unless stated otherwise.
This distinction is particularly important for stability diagnostics. Quantities protected by local symmetry satisfy Ward-type constraints at fixed kernel geometry, but the numerical location of an effective stability threshold may still depend on which kernel family is chosen to model the neural propagation sector.

Another useful aspect of the comparison with Abelian gauge theory is methodological. Once the MSRJD action is expanded around a quadratic background, the effective neural model admits diagrammatic corrections that can be organized in the same way as perturbative corrections in Euclidean field theory. The relevant diagrams are not to be interpreted as literal particle processes, but as bookkeeping devices for stochastic and finite-width corrections to the effective propagators and response functions.
In this sense, the gauge-theory analogy provides a disciplined perturbative language: self-energy-type insertions correspond to renormalization of effective propagators, vertex-type corrections encode modified couplings between fluctuation sectors, and Ward-type relations constrain the allowed longitudinal structure of these corrections. This is the sense in which gauge theory informs the neural analysis in the present work.

The structural comparison developed in this section has two practical consequences for the rest of the paper. First, it justifies importing symmetry-based reasoning, in particular Ward-type cancellations and gauge-covariant organization of fluctuations, into the stochastic neural setting. Second, it helps distinguish which claims are universal consequences of local \(U(1)\) covariance and which depend on additional modeling choices such as the effective potential, the kernel family, or the finite-width approximation scheme.
With these clarifications in place, the next section turns to the actual stability problem of interest: the two-replica linear-response analysis, the definition of the effective amplification factor, and the derivation of the edge-of-chaos criterion within the revised gauge-covariant stochastic model.

\section{Two-replica linear response, amplification factor, and edge-of-chaos criterion}
\label{sec:stability_edge_chaos}

We now address the central dynamical question: whether small perturbations decay, remain marginal, or grow under stochastic depth evolution. This is formulated as a linear-response stability problem within the gauge-covariant stochastic framework introduced in Secs.~\ref{sec:effective_model} and~\ref{sec:msrjd_observables}.
In this formulation, the stability criterion is defined entirely within the effective neural theory. The analogy with Abelian gauge theory serves only as structural guidance; all observables below are intrinsic to the stochastic dynamics.

To probe sensitivity to perturbations, we introduce two replicas \(a=1,2\) evolving under the same noise realization but with slightly different initial conditions:
\begin{eqnarray}
\Phi_a(t)=\big(\phi_a,\phi_a^\ast,W_{\mu,a}\big),\qquad
\tilde{\Phi}_a(t)=\big(\tilde{\phi}_a,\tilde{\phi}_a^\ast,\tilde{W}_{\mu,a}\big).
\end{eqnarray}
The replicated MSRJD action is
\begin{eqnarray}
S_{\rm 2rep}
=
S_{\rm MSRJD}[\Phi_1,\tilde{\Phi}_1]
+
S_{\rm MSRJD}[\Phi_2,\tilde{\Phi}_2],
\end{eqnarray}
with shared noise correlations,
\begin{eqnarray}
\langle \eta_1(x,t)\eta_2^\ast(x',t')\rangle
=
2\kappa_\phi\,\delta^{(d)}(x-x')\delta(t-t'),\\
\langle \xi_{\mu,1}(x,t)\xi_{\nu,2}(x',t')\rangle
=
2\kappa_W\,\delta_{\mu\nu}\delta^{(d)}(x-x')\delta(t-t').
\end{eqnarray}
Defining replica differences,
\begin{eqnarray}
\delta\phi:=\phi_1-\phi_2,\quad
\delta\phi^\ast:=\phi_1^\ast-\phi_2^\ast,\quad
\delta W_\mu:=W_{\mu,1}-W_{\mu,2},
\end{eqnarray}
one obtains deterministic linear evolution at leading order around a common stochastic background.

Let \(\Phi_c=(\Phi_1+\Phi_2)/2\) and \(\delta\Phi=\Phi_1-\Phi_2\). Linearization yields
\begin{eqnarray}
\partial_t \delta\Phi_A(x,t)
=
- \int d^d x'\;
\mathbb{H}_{AB}(x,x';\Phi_c)\,
\delta\Phi_B(x',t),
\end{eqnarray}
with Hessian kernel
\begin{eqnarray}
\mathbb{H}_{AB}(x,x';\Phi_c)
=
\frac{\delta^2 S_{\rm eff}}{\delta \Phi_A(x)\,\delta \Phi_B(x')}
\bigg|_{\Phi=\Phi_c}.
\end{eqnarray}
In momentum space,
\begin{eqnarray}
\partial_t \delta\Phi_A(p,t)
=
-\,\mathbb{H}_{AB}(p;\Phi_c)\,
\delta\Phi_B(p,t),
\end{eqnarray}
so stability is controlled by the spectrum of \(\mathbb{H}(p;\Phi_c)\).
Defining a quadratic norm,
\begin{eqnarray}
\|\delta\Phi(t)\|^2
=
\int \frac{d^dp}{(2\pi)^d}\,
\delta\Phi_A^\ast(p,t)\,
\mathbb{M}_{AB}(p)\,
\delta\Phi_B(p,t),
\end{eqnarray}
the maximal Lyapunov exponent is
\begin{eqnarray}
\lambda_{\max}
=
\lim_{t\to\infty}
\frac{1}{t}
\ln\frac{\|\delta\Phi(t)\|}{\|\delta\Phi(0)\|}.
\end{eqnarray}
The regimes are:
\begin{eqnarray}
\lambda_{\max}<0 \;\Rightarrow\; \text{stable},\quad
\lambda_{\max}=0 \;\Rightarrow\; \text{critical},\quad
\lambda_{\max}>0 \;\Rightarrow\; \text{unstable}.
\end{eqnarray}
The edge of chaos corresponds to \(\lambda_{\max}=0\).

Projecting onto the dominant mode \(p_\star\), we define
\begin{eqnarray}
\chi(g,\alpha;p_\star)
=
\frac{\mathcal{G}_{\rm dressed}(p_\star)}
{\mathcal{G}_{\rm bare}(p_\star)}.
\end{eqnarray}
The critical condition is
\begin{eqnarray}
\chi(g_c,\alpha;p_\star)=1.
\end{eqnarray}
Here \(\chi\) is always the full gain; corrections are treated separately.
At quadratic level,
\begin{eqnarray}
S_{\rm eff}=S^{(2)}+S^{(\rm int)},
\end{eqnarray}
leading to
\begin{eqnarray}
\chi_{\rm MF}
=
g^2
\int d\mu(\lambda)\,\rho(\lambda)\,f(\lambda;\alpha,m),
\end{eqnarray}
with criticality
\begin{eqnarray}
\chi_{\rm MF}(g_c,\alpha)=1.
\end{eqnarray}
Including interactions,
\begin{eqnarray}
\mathcal{G}_{\rm dressed}^{-1}(p)
=
\mathcal{G}_{\rm bare}^{-1}(p)-\Sigma_{\rm eff}(p),
\end{eqnarray}
and
\begin{eqnarray}
\chi
=
\chi_{\rm MF}
+
\delta\chi
+
\mathcal{O}(\cdots).
\end{eqnarray}
The critical condition remains \(\chi=1\) at all orders; only \(g_c\) may shift.
Local \(U(1)\) covariance imposes Ward-type constraints on \(\Sigma_{\rm eff}\). At the perturbative order considered,
\begin{eqnarray}
\chi
=
\chi_{\rm MF}
+
\delta\chi_{\rm reg},
\end{eqnarray}
with \(\delta\chi_{\rm reg}\) regular at criticality, so the marginal condition is not shifted.
Empirically, one measures
\begin{eqnarray}
\lambda_{\rm emp}
=
\frac{1}{L}
\Big\langle
\sum_{\ell=1}^{L}
\ln
\frac{\|\delta h_{\ell+1}\|}{\|\delta h_\ell\|}
\Big\rangle,
\end{eqnarray}
and compares \(\lambda_{\rm emp}=0\) with \(\chi=1\).
The stability analysis proceeds as follows: the two-replica construction defines a linearized problem; \(\lambda_{\max}\) provides the stability criterion; \(\chi\) summarizes the dominant mode; and the edge of chaos corresponds to \(\chi=1\) or \(\lambda_{\max}=0\). Symmetry-constrained corrections may deform propagation but do not shift criticality at the perturbative order considered.

\section{Numerical results}
\label{sec:numerical_results}

We present two numerical studies targeting complementary aspects of the framework. The first examines the onset of instability in finite-width multilayer perceptrons at initialization, comparing the empirical Lyapunov indicator with the mean-field amplification criterion. The second analyzes a linear stochastic effective model, used as a controlled setting to test the predicted finite-width spectral correction.

We consider fully connected networks of depth \(L\) and width \(N\), with Gaussian weights of variance \(\sigma_w^2/N\) and biases of variance \(\sigma_b^2\). For an activation \(\varphi\), the mean-field amplification factor is
\begin{eqnarray}
\chi_{\rm MF}(\sigma_w^2)
=
\sigma_w^2\,
\mathbb{E}\!\left[
\big(\varphi'(\sqrt{q^\ast}\,Z)\big)^2
\right],\qquad Z\sim\mathcal{N}(0,1),
\end{eqnarray}
where \(q^\ast\) is the variance fixed point. For ReLU, \(\chi_{\rm MF}=\sigma_w^2/2\), yielding the critical condition \(\chi_{\rm MF}=1\).
The empirical Lyapunov exponent is estimated as
\begin{eqnarray}
\lambda_{\rm emp}
=
\frac{1}{L}
\Big\langle
\sum_{\ell=1}^{L}
\log
\frac{\|\delta h_{\ell+1}\|}{\|\delta h_\ell\|}
\Big\rangle,
\end{eqnarray}
where \(\delta h_\ell\) is a small perturbation propagated through the same network realization.
We perform sweeps in \(\sigma_w^2\) for \(\tanh\) and ReLU activations with \(L=40\), \(N=200\), and zero bias variance.
\begin{figure}[ht]
    \centering
    \includegraphics[width=\linewidth]{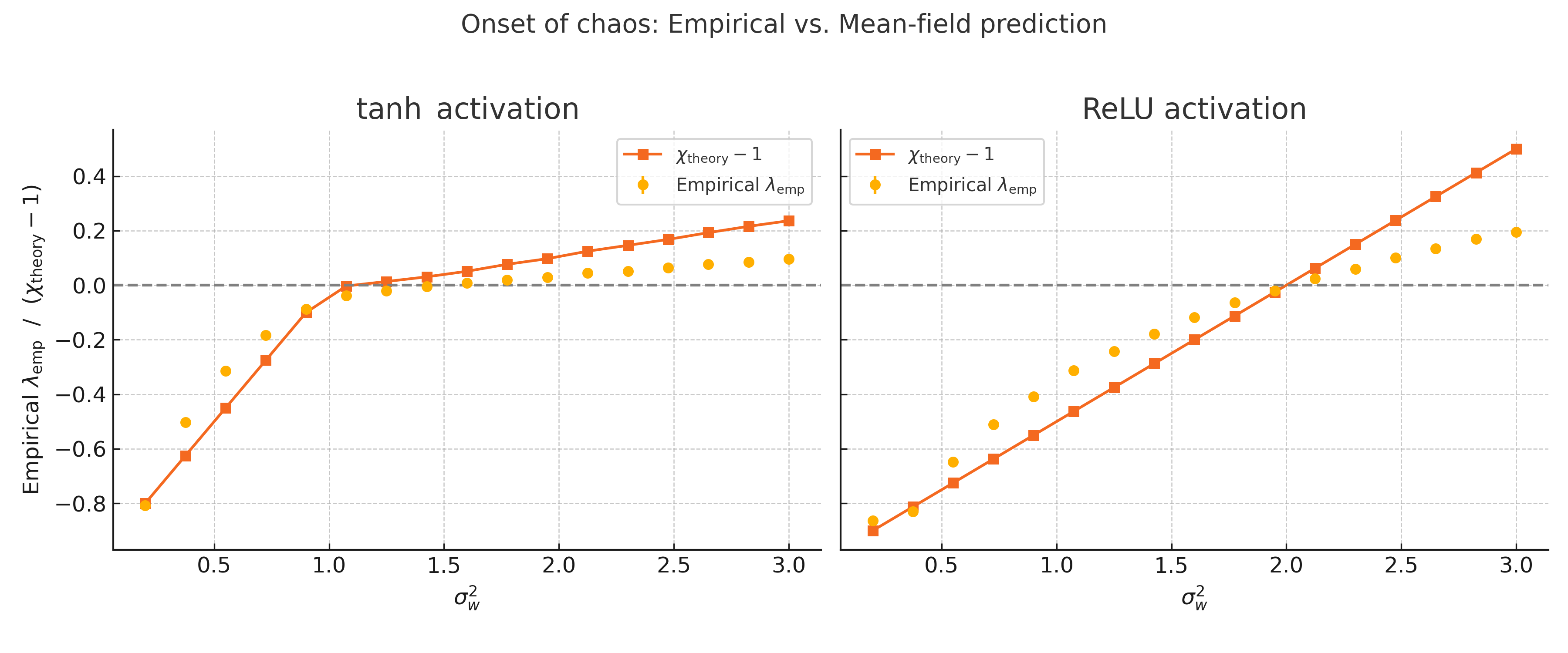}
    \caption{Empirical Lyapunov exponent per layer, \(\lambda_{\rm emp}\), and mean-field indicator, \(\chi_{\rm MF}-1\), as functions of \(\sigma_w^2\), for \(\tanh\) (left) and ReLU (right). The empirical transition occurs close to the mean-field threshold \(\chi_{\rm MF}=1\).}
    \label{fig:lambda_comparison}
\end{figure}
Figure~\ref{fig:lambda_comparison} shows that \(\lambda_{\rm emp}=0\) occurs near \(\chi_{\rm MF}=1\) for both activations, with small deviations due to finite depth and width.
We consider the linear stochastic dynamics
\begin{eqnarray}
\partial_t h_i(t)
=
-\,\gamma\,h_i(t)
+
\sum_{j=1}^{N}W_{ij}h_j(t)
+
\xi_i(t),
\end{eqnarray}
with
\begin{eqnarray}
W_{ij}\sim\mathcal{N}\!\left(0,\frac{\sigma_w^2}{N}\right),\qquad
\langle \xi_i(t)\xi_j(t')\rangle
=
2\kappa\,\delta_{ij}\delta(t-t').
\end{eqnarray}
The free correlator is
\begin{eqnarray}
c(\omega)=\frac{2\kappa}{\omega^2+\gamma^2},
\end{eqnarray}
and the theoretical spectrum
\begin{eqnarray}
X(\omega)
=
X^{(0)}(\omega)
+
\frac{\gamma T}{N}\,X^{(1)}(\omega),
\end{eqnarray}
with
\begin{eqnarray}
X^{(0)}(\omega)
=
c(\omega)\,
\frac{\omega^2+\gamma^2}{\omega^2+\gamma^2-\sigma_w^2},
\end{eqnarray}
\begin{eqnarray}
X^{(1)}(\omega)
=
\frac{\sigma_w^2}{2}\,
c(\omega)\,
\frac{\omega^2+\gamma^2}{(\omega^2+\gamma^2-\sigma_w^2)^2}.
\end{eqnarray}
We simulate the dynamics using Euler--Maruyama, compute the averaged signal
\begin{eqnarray}
s(t)=\frac{1}{N}\sum_{i=1}^{N}h_i(t),
\end{eqnarray}
and estimate its power spectrum via FFT, averaging over seeds. The theoretical curve is rescaled in the low-frequency region for shape comparison.
Parameters: \(\gamma=1\), \(\kappa=1\), \(dt=0.05\), \(L=2000\), warmup \(=400\), \(N=100\).
\begin{figure}[htbp]
  \centering
  \includegraphics[width=0.78\linewidth]{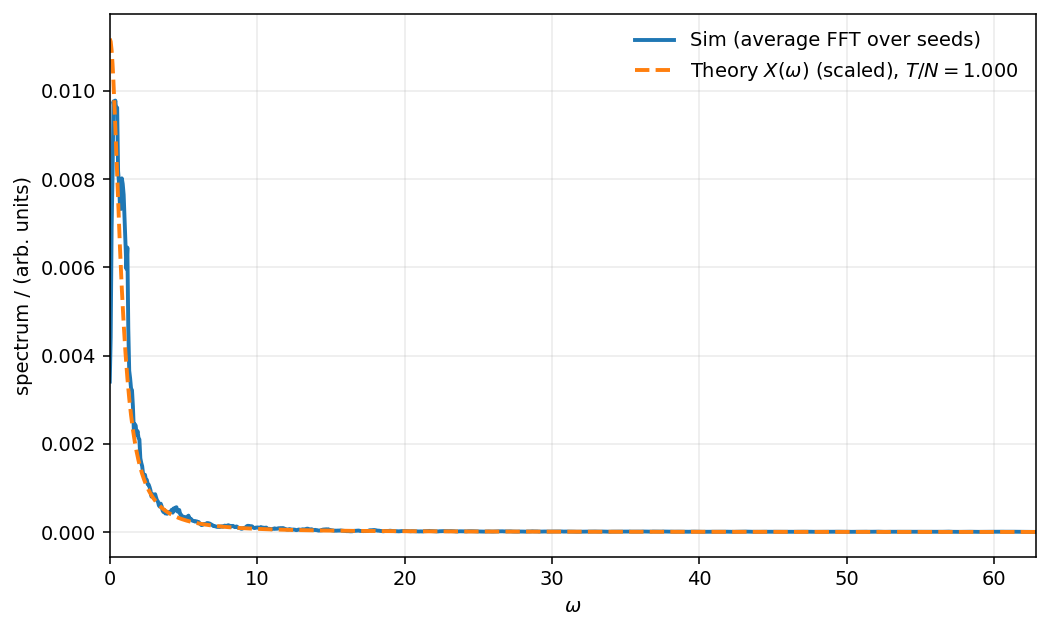}
  \caption{Simulated spectrum of the linear stochastic model compared with the theoretical prediction \(X(\omega)\). Agreement is observed in the low-frequency region.}
  \label{fig:theory_vs_sim_frequency}
\end{figure}
Figure~\ref{fig:theory_vs_sim_frequency} shows good agreement in the low-frequency regime, where the perturbative description is expected to hold.
Experiment A shows that the empirical instability threshold in finite-width networks is well captured by the mean-field amplification criterion. Experiment B shows that the leading finite-width correction reproduces the observed low-frequency spectral deformation in a controlled linear setting.

\section{Local symmetry, kernel geometry, and admissible observables}
\label{sec:local_symmetry_kernel}

A central point of the present framework is that local \(U(1)\) covariance constrains the effective stochastic dynamics, but does not imply that different propagation kernels correspond to the same neural model. This section clarifies that distinction and identifies the observables that are meaningful in the present formulation.
The effective action introduced in Sec.~\ref{sec:effective_model},
\begin{eqnarray}
S_{\rm eff}[\phi^\ast,\phi,W]
=
\int d^dx \,
\bigg[
(D_\mu\phi)^\ast (D_\mu\phi)
+ m^2 \phi^\ast\phi
+ U(\phi^\ast\phi)
+ \frac{1}{4}F_{\mu\nu}F_{\mu\nu}
+ \frac{1}{2\alpha}(\partial_\mu W_\mu)^2
\bigg],\nonumber\\
\end{eqnarray}
is built from gauge-covariant ingredients. At fixed \(\alpha\), it is invariant under
\begin{eqnarray}
\phi(x,t) \to e^{i\theta(x,t)}\phi(x,t),\\
\phi^\ast(x,t) \to e^{-i\theta(x,t)}\phi^\ast(x,t),\\
W_\mu(x,t) \to W_\mu(x,t)-\frac{1}{g}\partial_\mu\theta(x,t).
\end{eqnarray}
This symmetry expresses a redundancy in the phase-like parametrization of the effective feature field, compensated by a corresponding transformation of the connectivity field. Its role is structural: it restricts the admissible drift terms, constrains the longitudinal sector of the dressed kernels, and organizes perturbative corrections through Ward-type identities.

A fundamental distinction must be kept explicit. Local \(U(1)\) invariance at fixed model is not the same as varying the propagation kernel across different models. At fixed \(\alpha\), gauge-equivalent configurations describe the same effective theory, and properly defined gauge-invariant observables have the same expectation values. Examples include
\begin{eqnarray}
|\phi(x,t)|^2=\phi^\ast(x,t)\phi(x,t),
\end{eqnarray}
and
\begin{eqnarray}
F_{\mu\nu}(x,t)F_{\mu\nu}(x,t).
\end{eqnarray}
By contrast, changing \(\alpha\) changes the quadratic kernel of the connectivity sector. In ordinary Abelian gauge theory this is only a change of description. In the present neural interpretation, however, \(\alpha\) may also label different effective propagation geometries. Consequently, stability indicators, dressed kernels, and critical couplings need not remain numerically unchanged when \(\alpha\) is varied.

The correct statement is therefore the following: observables are invariant under local \(U(1)\) reparametrizations at fixed kernel geometry, but they are generally not invariant under changes of the kernel family itself.
The parameter \(\alpha\) enters through
\begin{eqnarray}
\frac{1}{2\alpha}(\partial_\mu W_\mu)^2,
\end{eqnarray}
which controls the longitudinal part of the connectivity-field propagator. In field-theory language this is the standard covariant-gauge sector. In the neural interpretation, the same term determines how connectivity fluctuations are weighted in the effective propagation kernel. For this reason, \(\alpha\) plays a dual role: it is a gauge parameter on the field-theory side and an effective kernel-geometry parameter on the neural side.

The meaningful observables of the effective model fall into three classes. First, there are strictly gauge-invariant local or integrated quantities, such as \(|\phi|^2\), \(F_{\mu\nu}F_{\mu\nu}\), and functionals built from them. Second, there are gauge-fixed correlation and response functions, such as the dressed kernels entering the amplification factor; these are meaningful only after the kernel geometry has been specified. Third, there are architecture-level effective observables, such as empirical Lyapunov indicators, correlation lengths, or spectral densities. These are not gauge-invariant in the strict field-theoretic sense, but they are the relevant diagnostics of the chosen effective propagation model and are precisely the quantities accessed in the numerical analysis.
For convenience, Table~\ref{tab:observables_dictionary_revised} summarizes the main quantities used in the present framework.
\begin{table}[ht]
\centering
\caption{Dictionary of observables and control parameters in the gauge-covariant stochastic framework.}
\label{tab:observables_dictionary_revised}
\begin{tabular}{|l|l|l|}
\hline
\textbf{Quantity} & \textbf{Field-theoretic role} & \textbf{Neural interpretation} \\
\hline
$\phi,\phi^\ast$ & Complex matter field & Coarse-grained feature or activation amplitude \\
\hline
$W_\mu$ & Abelian connection / gauge field & Effective connectivity or phase-transport field \\
\hline
$F_{\mu\nu}$ & Field strength & Connectivity curvature / nontrivial transport content \\
\hline
$g$ & Coupling strength & Effective interaction or propagation-strength parameter \\
\hline
$\alpha$ & Covariant-gauge parameter & Effective kernel-geometry parameter \\
\hline
$\kappa_\phi,\kappa_W$ & Noise amplitudes & Activation/update stochasticity scales \\
\hline
$\chi$ & Full dressed amplification factor & Stability indicator for perturbation growth \\
\hline
$\lambda_{\max}$ & Maximal linear-response growth rate & Edge-of-chaos / instability diagnostic \\
\hline
$C_{\phi\phi^\ast},\,C_{\mu\nu}$ & Two-point correlators & Feature and connectivity fluctuation propagation \\
\hline
$R_{\phi\phi^\ast},\,R_{\mu\nu}$ & Response functions & Sensitivity of propagation to perturbations \\
\hline
\end{tabular}
\end{table}
The present work is formulated as an effective continuum theory. It is therefore not intended to provide a unique microscopic realization for every neural architecture. Rather, it should be read as a coarse-grained description whose effective coordinate \(x\), dimension \(d\), and kernel geometry depend on the architecture under consideration.

For multilayer perceptrons, \(\phi(x,t)\) may be viewed as a continuum representation of feature amplitudes across neuron groups or channels, so that the relevant observables reduce to variance propagation, perturbation growth, and correlation decay. For convolutional or graph-based architectures, the same formalism suggests a local transport structure in which nearby spatial or graph modes are coupled through an effective connection field.
The purpose of the framework is not to prescribe a single explicit architecture, but to provide a symmetry-guided language for organizing stability analysis and effective kernel design across different model families.

This has an important consequence for the interpretation of the results. When the manuscript states that the critical condition is protected by symmetry at a given perturbative order, this statement is always meant at fixed kernel geometry. It does not imply that different kernel families, or different choices of \(\alpha\) interpreted as different neural propagation geometries, share the same numerical critical point.
The claim is therefore more precise than in the original version: local \(U(1)\) covariance constrains the form of the dressed kernels and can prevent specific perturbative corrections from shifting the marginality condition within a chosen model class, but it does not remove genuine model dependence associated with the choice of effective kernel geometry.


\section{Conclusion}
\label{sec:conclusion}

We developed a gauge-covariant stochastic field framework for the analysis of stability and finite-width effects in deep neural systems. The formulation is built entirely from classical commuting fields: a complex matter field representing coarse-grained feature amplitudes, a real Abelian connection field encoding effective connectivity, and a stochastic depth variable governing noisy propagation. This defines a mathematically consistent effective theory in which local \(U(1)\) covariance, response theory, and perturbative corrections can be treated within a unified functional framework.

The relation to gauge theory is structural rather than literal. The framework does not represent a physical realization of quantum electrodynamics, but instead imports its organizing principles—local phase covariance, covariant derivatives, gauge fixing, and Ward-type constraints—to structure the analysis of neural dynamics.
Within this setting, stability is formulated through a two-replica linear-response construction. This allows a direct definition of the maximal Lyapunov exponent and of the amplification factor within the effective theory. The edge-of-chaos condition is identified with the marginality condition \(\lambda_{\max}=0\), equivalently \(\chi=1\), where \(\chi\) denotes the full dressed amplification factor in the dominant fluctuation channel.

The numerical results support this interpretation. In finite-width multilayer perceptrons, the empirical instability threshold closely follows the mean-field amplification criterion. In a controlled linear stochastic effective model, the leading finite-width correction reproduces the observed low-frequency spectral deformation, consistent with the dressed-kernel picture.
The role of symmetry is also clarified. Local \(U(1)\) covariance constrains the structure of dressed propagators and response functions at fixed kernel geometry, but does not eliminate model dependence associated with different kernel families. In particular, the gauge parameter \(\alpha\) can be interpreted as labeling distinct effective propagation geometries, so critical points need not be invariant under its variation. Symmetry protection of the critical condition therefore holds only within a fixed kernel class.

Therefore, these results support a consistent picture: neural stability can be described by a gauge-covariant effective field theory in which the edge of chaos emerges as a symmetry-constrained marginality condition, while finite-width effects appear as perturbative deformations of propagation kernels. This provides a principled alternative to purely heuristic initialization criteria.
Several directions remain open. On the theoretical side, extending the analysis to nonlinear dressed sectors and higher-order corrections would clarify the robustness of the critical surface. On the modeling side, it remains to be understood how different architectural classes—such as convolutional, graph-based, or equivariant networks—map onto distinct effective kernel geometries.

The main claim is thus precise: local gauge-covariant structure provides a useful organizing principle for stochastic neural stability, yielding controlled diagnostics of marginality and finite-width effects without requiring a literal identification with quantum field theory.

\section*{Funding}

This research received no external funding.

\section*{Data availability}

All data generated or analysed during this study are included in this published article.

\appendix

\section{MSRJD representation of the gauge-covariant stochastic model}
\label{app:msrjd_derivation}

In this appendix we derive the Martin--Siggia--Rose--Janssen--de~Dominicis (MSRJD) representation of the gauge-covariant stochastic model introduced in the main text. The construction follows directly from It\^o Langevin dynamics for commuting fields with Gaussian noise.
The model is defined in terms of a complex matter field \(\phi(x,t)\), its conjugate \(\phi^\ast(x,t)\), and a real Abelian connection field \(W_\mu(x,t)\), evolving according to
\begin{eqnarray}
\partial_t \phi(x,t)
=
-\,\frac{\delta S_{\rm eff}}{\delta \phi^\ast(x,t)}
+\eta(x,t),
\label{eq:app_langevin_phi}
\\
\partial_t \phi^\ast(x,t)
=
-\,\frac{\delta S_{\rm eff}}{\delta \phi(x,t)}
+\eta^\ast(x,t),
\label{eq:app_langevin_phistar}
\\
\partial_t W_\mu(x,t)
=
-\,\frac{\delta S_{\rm eff}}{\delta W_\mu(x,t)}
+\xi_\mu(x,t),
\label{eq:app_langevin_W}
\end{eqnarray}
with effective action
\begin{eqnarray}
S_{\rm eff}[\phi^\ast,\phi,W]
=
\int d^dx\,
\bigg[
(D_\mu\phi)^\ast (D_\mu\phi)
+ m^2 \phi^\ast\phi
+ U(\phi^\ast\phi)
+ \frac{1}{4}F_{\mu\nu}F_{\mu\nu}
+ \frac{1}{2\alpha}(\partial_\mu W_\mu)^2
\bigg].\nonumber\\
\end{eqnarray}

The noises are Gaussian and white,
\begin{eqnarray}
\langle \eta(x,t)\eta^\ast(x',t')\rangle
=
2\kappa_\phi\,\delta^{(d)}(x-x')\delta(t-t'),
\\
\langle \xi_\mu(x,t)\xi_\nu(x',t')\rangle
=
2\kappa_W\,\delta_{\mu\nu}\delta^{(d)}(x-x')\delta(t-t').
\end{eqnarray}

The noise probability measure is
\begin{eqnarray}
\mathcal{P}[\eta,\eta^\ast,\xi]
\propto
\exp\Bigg[
-\frac{1}{2\kappa_\phi}
\int dt\,d^dx\,\eta^\ast\eta
-\frac{1}{4\kappa_W}
\int dt\,d^dx\,\xi_\mu\xi_\mu
\Bigg].
\end{eqnarray}

Eliminating the noises using the Langevin equations yields the stochastic path weight
\begin{eqnarray}
\mathcal{P}[\phi,\phi^\ast,W]
\propto
\exp\Bigg[
-\frac{1}{2\kappa_\phi}
\int dt\,d^dx\,
\Big(
\partial_t\phi+\frac{\delta S_{\rm eff}}{\delta\phi^\ast}
\Big)
\Big(
\partial_t\phi^\ast+\frac{\delta S_{\rm eff}}{\delta\phi}
\Big)
\nonumber\\
-\frac{1}{4\kappa_W}
\int dt\,d^dx\,
\Big(
\partial_tW_\mu+\frac{\delta S_{\rm eff}}{\delta W_\mu}
\Big)^2
\Bigg].
\end{eqnarray}

For functional and perturbative analysis, it is convenient to rewrite the theory in first-order MSRJD form. This is achieved by enforcing the Langevin equations with functional delta distributions and introducing response fields \(\tilde{\phi}\), \(\tilde{\phi}^\ast\), and \(\tilde{W}_\mu\). In the It\^o discretization, the associated Jacobian is field independent and can be absorbed into the normalization.

Integrating out the noises, one obtains the generating functional
\begin{eqnarray}
Z
=
\int
\mathcal{D}\phi\,\mathcal{D}\phi^\ast\,\mathcal{D}W\,
\mathcal{D}\tilde{\phi}\,\mathcal{D}\tilde{\phi}^\ast\,\mathcal{D}\tilde{W}\;
e^{-S_{\rm MSRJD}},
\end{eqnarray}
with MSRJD action
\begin{eqnarray}
S_{\rm MSRJD}
=
\int dt\,d^dx\,
\Bigg\{
\tilde{\phi}^\ast
\Big[
\partial_t\phi+\frac{\delta S_{\rm eff}}{\delta\phi^\ast}
\Big]
+
\tilde{\phi}
\Big[
\partial_t\phi^\ast+\frac{\delta S_{\rm eff}}{\delta\phi}
\Big]
\nonumber\\
+
\tilde{W}_\mu
\Big[
\partial_tW_\mu+\frac{\delta S_{\rm eff}}{\delta W_\mu}
\Big]
-\kappa_\phi\,\tilde{\phi}^\ast\tilde{\phi}
-\kappa_W\,\tilde{W}_\mu\tilde{W}_\mu
\Bigg\}.
\label{eq:app_MSRJD_final}
\end{eqnarray}

Expanding around a background configuration,
\begin{eqnarray}
\phi=\phi_c+\delta\phi,\qquad
\phi^\ast=\phi_c^\ast+\delta\phi^\ast,\qquad
W_\mu=W_{\mu,c}+\delta W_\mu,
\end{eqnarray}
the quadratic part of \(S_{\rm MSRJD}\) defines the bare response functions and correlators. The response sector is first order in \(\partial_t\), while correlations are controlled by \(\kappa_\phi\) and \(\kappa_W\). Higher-order terms generate perturbative corrections, which in the neural interpretation correspond to finite-width and interaction effects.

The MSRJD action inherits the local \(U(1)\) structure of \(S_{\rm eff}\). With appropriate transformation rules for the response fields, the functional maintains the same covariance properties as the underlying stochastic equations.

This construction establishes that the model defines a consistent stochastic field theory with a well-defined functional representation, providing the basis for the response analysis and perturbative framework used in the main text.

\section{Local symmetry and Ward-type identity in the effective stochastic model}
\label{app:ward_identity}

In this appendix we derive the local symmetry statement and the associated Ward-type identity for the effective stochastic model. The goal is to make explicit how local \(U(1)\) covariance constrains correlation functions and dressed kernels.

The fields \(\phi(x,t)\), \(\phi^\ast(x,t)\), and \(W_\mu(x,t)\) transform under a local phase rotation as
\begin{eqnarray}
\phi(x,t) \to e^{i\theta(x,t)}\phi(x,t),
\\
\phi^\ast(x,t) \to e^{-i\theta(x,t)}\phi^\ast(x,t),
\\
W_\mu(x,t) \to W_\mu(x,t)-\frac{1}{g}\partial_\mu\theta(x,t).
\label{eq:ward_transf_W}
\end{eqnarray}
The covariant derivative and field strength are
\begin{eqnarray}
D_\mu\phi=(\partial_\mu+i g W_\mu)\phi,
\qquad
(D_\mu\phi)^\ast=(\partial_\mu-i g W_\mu)\phi^\ast,
\\
F_{\mu\nu}=\partial_\mu W_\nu-\partial_\nu W_\mu.
\end{eqnarray}
Under these transformations,
\begin{eqnarray}
D_\mu\phi \to e^{i\theta}D_\mu\phi,
\qquad
(D_\mu\phi)^\ast \to e^{-i\theta}(D_\mu\phi)^\ast,
\qquad
F_{\mu\nu}\to F_{\mu\nu},
\end{eqnarray}
so the combinations \((D_\mu\phi)^\ast(D_\mu\phi)\), \(\phi^\ast\phi\), \(U(\phi^\ast\phi)\), and \(F_{\mu\nu}F_{\mu\nu}\) are invariant.
The effective action is
\begin{eqnarray}
S_{\rm eff}[\phi^\ast,\phi,W]
=
\int d^dx\,
\bigg[
(D_\mu\phi)^\ast(D_\mu\phi)
+m^2\phi^\ast\phi
+U(\phi^\ast\phi)
+\frac{1}{4}F_{\mu\nu}F_{\mu\nu}
+\frac{1}{2\alpha}(\partial_\mu W_\mu)^2
\bigg].\nonumber\\
\label{eq:ward_Seff}
\end{eqnarray}
The first four terms are invariant, while the last term implements covariant gauge fixing. The symmetry is therefore realized in the standard gauge-fixed sense, and identities derived below are understood at fixed \(\alpha\).
We introduce the generating functional
\begin{eqnarray}
Z[J,J^\ast,J_\mu]
=
\int \mathcal{D}\phi\,\mathcal{D}\phi^\ast\,\mathcal{D}W\;
\exp\Bigg[
- S_{\rm eff}
+\int d^dx\,
\big(
J^\ast\phi+J\phi^\ast+J_\mu W_\mu
\big)
\Bigg].
\label{eq:ward_Z}
\end{eqnarray}
Consider an infinitesimal transformation \(\theta=\varepsilon\,\vartheta\). The variations are
\begin{eqnarray}
\delta\phi=i\varepsilon\,\vartheta\,\phi,
\qquad
\delta\phi^\ast=-i\varepsilon\,\vartheta\,\phi^\ast,
\qquad
\delta W_\mu=-\frac{\varepsilon}{g}\partial_\mu\vartheta.
\end{eqnarray}
Since the functional measure is invariant, the variation of \(Z\) vanishes. Only the source terms and the gauge-fixing term contribute. After integration by parts, one obtains
\begin{eqnarray}
0
=
\Bigg\langle
iJ^\ast\phi
-iJ\phi^\ast
+\frac{1}{g}\partial_\mu J_\mu
-\frac{1}{\alpha g}\,\partial^2(\partial_\mu W_\mu)
\Bigg\rangle_{J}.
\label{eq:ward_identity_basic}
\end{eqnarray}
This is the Ward-type identity of the gauge-fixed theory. In the absence of sources,
\begin{eqnarray}
\Big\langle
\partial^2(\partial_\mu W_\mu)
\Big\rangle = 0,
\end{eqnarray}
reflecting the fact that the longitudinal sector is fixed by the gauge condition.
Differentiation with respect to sources generates relations among correlation functions. In particular, for the two-point connectivity kernel,
\begin{eqnarray}
\Gamma_{\mu\nu}(p)
=
\Gamma_T(p)\,
\left(
\delta_{\mu\nu}-\frac{p_\mu p_\nu}{p^2}
\right)
+
\Gamma_L(p)\,
\frac{p_\mu p_\nu}{p^2},
\end{eqnarray}
the identity constrains the longitudinal component and relates it to the matter sector.
As a consequence, perturbative corrections are not independent: longitudinal structures are restricted, and certain deformations cannot modify the marginal stability condition associated with the dominant fluctuation channel. This is the sense in which symmetry constrains the stability analysis.

The same construction applies to the MSRJD functional. Assigning covariant transformation rules to the response fields yields identical identities for response and mixed correlators.
These Ward-type relations provide the functional constraints underlying the perturbative analysis of the main text.

\section{Leading perturbative correction to the amplification kernel}
\label{app:oneloop}

In this appendix we formulate the leading perturbative correction to the amplification kernel directly within the effective stochastic theory. The purpose is to identify the first dressed-kernel correction relevant to marginality and to state precisely in what sense the critical point is unchanged at the perturbative order considered.
Starting from the MSRJD functional, we expand around a homogeneous stationary background and write
\begin{eqnarray}
S_{\rm MSRJD}=S^{(2)}+S^{(\rm int)},
\end{eqnarray}
where \(S^{(2)}\) contains the quadratic fluctuation terms and \(S^{(\rm int)}\) the interaction terms generated by the gauge-covariant matter sector, the potential \(U(\phi^\ast\phi)\), and the nonlinear drift contributions.
The quadratic sector defines the bare inverse kernel
\begin{eqnarray}
\mathcal{K}_0(p,\omega),
\end{eqnarray}
whose inverse gives the propagator in the channel relevant for perturbation growth. Let \(p_\star\) denote the mode that first becomes marginal as the control parameter \(g\) is varied. The mean-field critical point \(g_c^{(0)}\) is determined by
\begin{eqnarray}
\mathcal{K}_0(p_\star,0;g_c^{(0)})=0,
\label{eq:bare_marginality}
\end{eqnarray}
or equivalently
\begin{eqnarray}
\chi_{\rm MF}(g_c^{(0)},\alpha;p_\star)=1.
\end{eqnarray}
Interactions dress the kernel according to
\begin{eqnarray}
\mathcal{K}(p,\omega;g)
=
\mathcal{K}_0(p,\omega;g)-\Sigma_{\rm eff}(p,\omega;g),
\label{eq:dressed_kernel_app}
\end{eqnarray}
where \(\Sigma_{\rm eff}\) is the leading self-energy correction in the dominant channel.
Near the mean-field critical point and at low frequency,
\begin{eqnarray}
\mathcal{K}(p_\star,0;g)
=
A\,(g-g_c^{(0)})
-\Sigma_{\rm eff}(p_\star,0;g_c^{(0)})
+\mathcal{O}\!\left((g-g_c^{(0)})^2\right)
+\mathcal{O}\!\left((g-g_c^{(0)})\,\Sigma_{\rm eff}\right),
\label{eq:kernel_expansion_critical}
\end{eqnarray}
with
\begin{eqnarray}
A=
\left.
\frac{\partial \mathcal{K}_0(p_\star,0;g)}{\partial g}
\right|_{g=g_c^{(0)}}.
\end{eqnarray}
The dressed critical point \(g_c\) is defined by
\begin{eqnarray}
\mathcal{K}(p_\star,0;g_c)=0.
\label{eq:shifted_critical_condition}
\end{eqnarray}
To first order,
\begin{eqnarray}
g_c-g_c^{(0)}
=
\frac{\Sigma_{\rm eff}(p_\star,0;g_c^{(0)})}{A}
+\mathcal{O}(\Sigma_{\rm eff}^2).
\label{eq:shifted_gc_general}
\end{eqnarray}
Thus the critical point shifts only if the dressed correction has a nonvanishing projection onto the critical mode at \(p_\star\).
In the present framework, the non-shift statement is meant in the following restricted sense: at the perturbative order considered, the leading correction does not contribute in the critical projection. This may occur because the correction is regular in a noncritical sector, or because longitudinal contributions are constrained by the Ward-type identity of Appendix~\ref{app:ward_identity} and therefore vanish when projected onto the marginal channel at fixed kernel geometry. The relevant statement is
\begin{eqnarray}
\Sigma_{\rm eff}(p_\star,0;g_c^{(0)})=0
\qquad
\text{in the critical projection at the perturbative order considered}.
\label{eq:no_shift_projection}
\end{eqnarray}
Substituting Eq.~\eqref{eq:no_shift_projection} into Eq.~\eqref{eq:shifted_gc_general} gives
\begin{eqnarray}
g_c=g_c^{(0)}+\mathcal{O}(\Sigma_{\rm eff}^2).
\label{eq:no_shift_gc}
\end{eqnarray}
This is the precise content of the non-shift statement: the dressed kernel is modified, but the marginality condition is not displaced at the perturbative order retained.
The same conclusion can be expressed in terms of the full amplification factor \(\chi\). Writing
\begin{eqnarray}
\chi(g,\alpha;p_\star)
=
\chi_{\rm MF}(g,\alpha;p_\star)
+
\delta\chi(g,\alpha;p_\star),
\label{eq:chi_split_appendix}
\end{eqnarray}
the critical condition remains
\begin{eqnarray}
\chi(g_c,\alpha;p_\star)=1.
\label{eq:chi_critical_appendix}
\end{eqnarray}
Expanding around \(g_c^{(0)}\),
\begin{eqnarray}
1
=
\chi_{\rm MF}(g_c^{(0)},\alpha;p_\star)
+
(g_c-g_c^{(0)})
\left.
\frac{\partial\chi_{\rm MF}}{\partial g}
\right|_{g_c^{(0)}}
+
\delta\chi(g_c^{(0)},\alpha;p_\star)
+\cdots.
\end{eqnarray}
Since \(\chi_{\rm MF}(g_c^{(0)},\alpha;p_\star)=1\),
\begin{eqnarray}
g_c-g_c^{(0)}
=
-
\frac{
\delta\chi(g_c^{(0)},\alpha;p_\star)
}{
\left.
\frac{\partial\chi_{\rm MF}}{\partial g}
\right|_{g_c^{(0)}}
}
+\cdots.
\label{eq:shift_gc_chi}
\end{eqnarray}
Hence the critical point is unchanged at this order precisely when
\begin{eqnarray}
\delta\chi(g_c^{(0)},\alpha;p_\star)=0
\end{eqnarray}
in the critical channel.
This also resolves the notation issue present in the earlier version: \(\chi\) always denotes the full gain, while \(\delta\chi\) denotes only the perturbative correction. The critical point is always defined by the full condition \(\chi=1\).
For the linear stochastic effective model used in the numerical section,
\begin{eqnarray}
\partial_t h_i(t)
=
-\,\gamma\,h_i(t)
+
\sum_{j=1}^{N}W_{ij}h_j(t)
+
\xi_i(t),
\end{eqnarray}
the dressed spectrum is written as
\begin{eqnarray}
X(\omega)
=
X^{(0)}(\omega)
+
\frac{\gamma T}{N}\,X^{(1)}(\omega).
\end{eqnarray}
In this sector, \(X^{(1)}\) deforms the spectral shape but does not generate a new zero of the inverse kernel at the perturbative order retained. Equivalently, the correction is regular at the mean-field marginal point and leaves the critical projection unchanged. This justifies comparing spectral deformation numerically without introducing a shifted critical parameter at the same order.

This conclusion is restricted to the linear effective sector and to the perturbative order retained. Higher-order corrections, different kernel geometries, or different truncation schemes may shift the critical surface.
The role of this appendix is therefore to replace the original fermionic one-loop argument by a perturbative statement internal to the revised effective theory: the first dressed-kernel correction may renormalize amplitudes and spectral shapes, yet the edge-of-chaos condition remains unchanged at that order whenever the correction has vanishing projection onto the critical mode.


\section{Finite-width expansion in the linear stochastic effective sector}
\label{app:finite_width_linear}

In this appendix we summarize the finite-width expansion for the linear stochastic effective model used in the numerical section. The goal is to clarify the origin of the \(T/N\) scaling, the structure of the leading correction, and its effect on the spectral kernel.
We consider the stochastic dynamics
\begin{eqnarray}
\partial_t h_i(t)
=
-\,\gamma\,h_i(t)
+
\sum_{j=1}^{N}W_{ij}h_j(t)
+
\xi_i(t),
\label{eq:app_linear_model}
\end{eqnarray}
with random couplings
\begin{eqnarray}
W_{ij}\sim \mathcal{N}\!\left(0,\frac{\sigma_w^2}{N}\right),
\end{eqnarray}
and Gaussian white noise
\begin{eqnarray}
\langle \xi_i(t)\xi_j(t')\rangle
=
2\kappa\,\delta_{ij}\delta(t-t').
\end{eqnarray}
In the absence of couplings, the process reduces to an Ornstein--Uhlenbeck dynamics with frequency-space correlator
\begin{eqnarray}
c(\omega)=\frac{2\kappa}{\omega^2+\gamma^2}.
\label{eq:app_free_correlator}
\end{eqnarray}
At large \(N\), averaging over the random couplings produces a dressed correlator \(X(\omega)\). The leading (infinite-width) contribution is
\begin{eqnarray}
X^{(0)}(\omega)
=
c(\omega)\,
\frac{\omega^2+\gamma^2}{\omega^2+\gamma^2-\sigma_w^2},
\label{eq:app_X0}
\end{eqnarray}
which plays the role of the mean-field propagator.
Finite-width corrections arise from contractions of the random couplings,
\begin{eqnarray}
\langle W_{ij}W_{kl}\rangle
=
\frac{\sigma_w^2}{N}\,\delta_{ik}\delta_{jl}.
\label{eq:app_W_contraction}
\end{eqnarray}
Each contraction contributes a factor \(1/N\), while sums over internal indices contribute factors of \(N\). In the stochastic setting, time integrals introduce an additional scale: the observation window \(T\). The leading correction contains one unconstrained time integral, producing a factor proportional to \(T\). As a result, the effective expansion parameter is
\begin{eqnarray}
\frac{T}{N}.
\end{eqnarray}
To first subleading order, the spectrum is written as
\begin{eqnarray}
X(\omega)
=
X^{(0)}(\omega)
+
\frac{\gamma T}{N}\,X^{(1)}(\omega)
+
\mathcal{O}\!\left((T/N)^2\right),
\label{eq:app_Xexpansion}
\end{eqnarray}
with
\begin{eqnarray}
X^{(1)}(\omega)
=
\frac{\sigma_w^2}{2}\,
c(\omega)\,
\frac{\omega^2+\gamma^2}{(\omega^2+\gamma^2-\sigma_w^2)^2}.
\label{eq:app_X1}
\end{eqnarray}

The resulting expression,
\begin{eqnarray}
X(\omega)
=
c(\omega)\,
\frac{\omega^2+\gamma^2}{\omega^2+\gamma^2-\sigma_w^2}
+
\frac{\gamma T}{N}\,
\frac{\sigma_w^2}{2}\,
c(\omega)\,
\frac{\omega^2+\gamma^2}{(\omega^2+\gamma^2-\sigma_w^2)^2}
+
\mathcal{O}\!\left((T/N)^2\right),
\label{eq:app_final_Xomega}
\end{eqnarray}
is the form used in the numerical comparison.
The leading term \(X^{(0)}(\omega)\) determines the marginal condition through the vanishing of the inverse denominator. The correction \(X^{(1)}(\omega)\) modifies the spectral shape but remains regular at the mean-field marginal point. Therefore, at this perturbative order, it does not introduce a new zero of the inverse kernel and does not shift the marginality condition.

This statement is restricted to the linear effective sector and to the order retained in the expansion. Higher-order terms or nonlinear effects may modify the critical point.
In the simulations, one measures the averaged signal
\begin{eqnarray}
s(t)=\frac{1}{N}\sum_{i=1}^{N}h_i(t),
\end{eqnarray}
computes its FFT spectrum, and compares the result with Eq.~\eqref{eq:app_final_Xomega}. Since the normalization depends on discretization and FFT conventions, the comparison is performed at the level of spectral shape.
The agreement observed in the low-frequency regime supports the use of Eq.~\eqref{eq:app_final_Xomega} as a leading-order description of finite-width spectral deformation in this linear stochastic setting.

\bibliographystyle{unsrtnat}
\bibliography{biblio}

\end{document}